\documentclass[pdflatex,sn-mathphys-num]{sn-jnl}

\usepackage{graphicx}%
\usepackage{multirow}%
\usepackage{amsmath,amssymb,amsfonts}%
\usepackage{amsmath,amssymb,graphicx,bbold}
\usepackage{amsthm}%
\usepackage{mathrsfs}%
\usepackage{physics}
\usepackage{float}
\usepackage{bm}
\usepackage{xcolor}%
\usepackage{textcomp}%
\usepackage{manyfoot}%
\usepackage{booktabs}%
\usepackage{listings}%

\newcommand{\ii}{\mathrm{i}}
\newcommand{\e}{\mathrm{e}}

\theoremstyle{thmstyleone}%
\theoremstyle{thmstyletwo}%

\theoremstyle{thmstylethree}%

\begin{document}

\title[Experimental investigation of the role of spatial correlations in
optical integration with heralded single photons]{Experimental investigation of the role of spatial correlations in
optical integration with heralded single photons}


\author*[1]{\fnm{L.} \sur{Marques Fagundes}}

\author[1]{\fnm{R. C.} \sur{Souza Pimenta}}
\equalcont{These authors contributed equally to this work.}

\author[2]{\fnm{M. H.} \sur{Magiotto}}
\equalcont{These authors contributed equally to this work.}

\author[2]{\fnm{R. M.} \sur{Gomes}}

\author[1]{\fnm{E. I.} \sur{Duzzioni}}

\author[1]{\fnm{R.} \sur{Medeiros de Araújo}}

\author[1]{\fnm{P. H. } \sur{Souto Ribeiro}}

\affil*[1]{\orgdiv{Departamento}, \orgname{Universidade Federal de Santa Catarina}, \orgaddress{\street{Rua Roberto Sampaio Gonzaga}, \city{Florianópolis}, \postcode{88040-900}, \state{Santa Catarina}, \country{Brasil}}}

\affil[2]{\orgdiv{Instituto de F\'{i}sica}, \orgname{Universidade Federal de Goi\'as}, \orgaddress{\street{Avenida Esperança}, \city{Goiânia}, \postcode{74690-900}, \state{Goiás}, \country{Brasil}}}


\abstract{In this work, we demonstrate optical integration using heralded single photons and explore the influence of spatial correlations between photons on this process. Specifically, we experimentally harness the transverse spatial degrees of freedom of light within an optical processing framework based on heralded single photons. The integration is performed over binary phase patterns encoded via a phase-only spatial light modulator, with polarization serving as an auxiliary degree of freedom.
Our findings reveal a distinct contrast in how spatial correlations affect image analysis: spatially uncorrelated photons are more effective at capturing the global features of an image encoded in the modulator, whereas spatially correlated photons exhibit enhanced sensitivity to local image details. Importantly, the optical integration scheme presented here bears a strong conceptual and operational resemblance to the DQC1 (Deterministic Quantum Computation with One Qubit) model. This connection underscores the potential of our approach for quantum-enhanced information processing, even in regimes where entanglement is minimal or absent.}




\maketitle

\section{Introduction}\label{sec1}
The rapid growth of computational demands in fields such as artificial intelligence, cryptography, and large-scale simulations has spurred the search for novel computing paradigms that transcend the limitations of classical electronic systems \cite{gill2022quantum}. Optical computing, which uses photons as information carriers, has emerged as a promising alternative due to its inherent advantages in speed, parallelism, and energy efficiency \cite{Miller2010,MAGIOTTO2025112137,hu2024diffractive, mcmahon2023physics, hengeveld2021optical}. Among the various approaches to optical computing, parallel architectures that exploit the wave nature of light have received significant attention for their ability to perform multiple operations simultaneously \cite{Caulfield2010,mengu2023diffractive,kulce2021all,huang2022orbital,nape2024optical,bezuidenhout2024variational}. The generation of single photons via spontaneous parametric down-conversion (SPDC) has further expanded the potential of optical computing by enabling the integration of quantum states of light into computational frameworks \cite{Kwiat1995, jennewein2011single, walther2005experimental}. SPDC provides a robust source of single photons with high purity and entanglement capabilities, making it an ideal candidate for realizing parallel optical computing systems with quantum-enhanced functionalities \cite{barz2010heralded, pittman2003heralded} as explored in several tasks over recent years, including the creation of quantum algorithms \cite{qu2024experimental, weng2024implementation, rojas2024non}, quantum cryptography and quantum key distribution \cite{hreibi2025entanglement, kravtsov2023security,kim2025efficient,karthik2025noise}. Also, the spatial degree of freedom is utilized in ghost imaging \cite{pearce2023practical, defienne2024advances, ma2025quantum}, the creation of high-dimensional Hilbert space \cite{nape2023quantum, d2016entangled}, and other applications \cite{lin2025observation, zhang2024entanglement, pimenta2024photonic}.

In this work, we present an optical computing architecture that leverages single photons generated through SPDC to perform computational tasks. By exploiting the quantum properties of single photons and, correlations between the signal and idler photons, it's possible to achieve the simultaneous execution of multiple operations, offering a significant advantage over classical computing approaches. For example, the integration of SPDC-based single photons into optical circuits allows for the realization of highly parallelized computations, with applications that include quantum simulation, optimization, and classical data processing \cite{Knill2001,AspuruGuzik2012}.

Here, we experimentally investigate an optical processing scheme based on the use of heralded single photons produced in SPDC, phase modulation with a spatial light modulator (SLM), and polarization-assisted phase-to-amplitude conversion. We perform optical integration of binary functions and compare the performance of the system for two types of SPDC spatial correlation engineering.

\section{Experiment}\label{exp}

\begin{figure*}[t!]
	\includegraphics[width=0.9\textwidth]{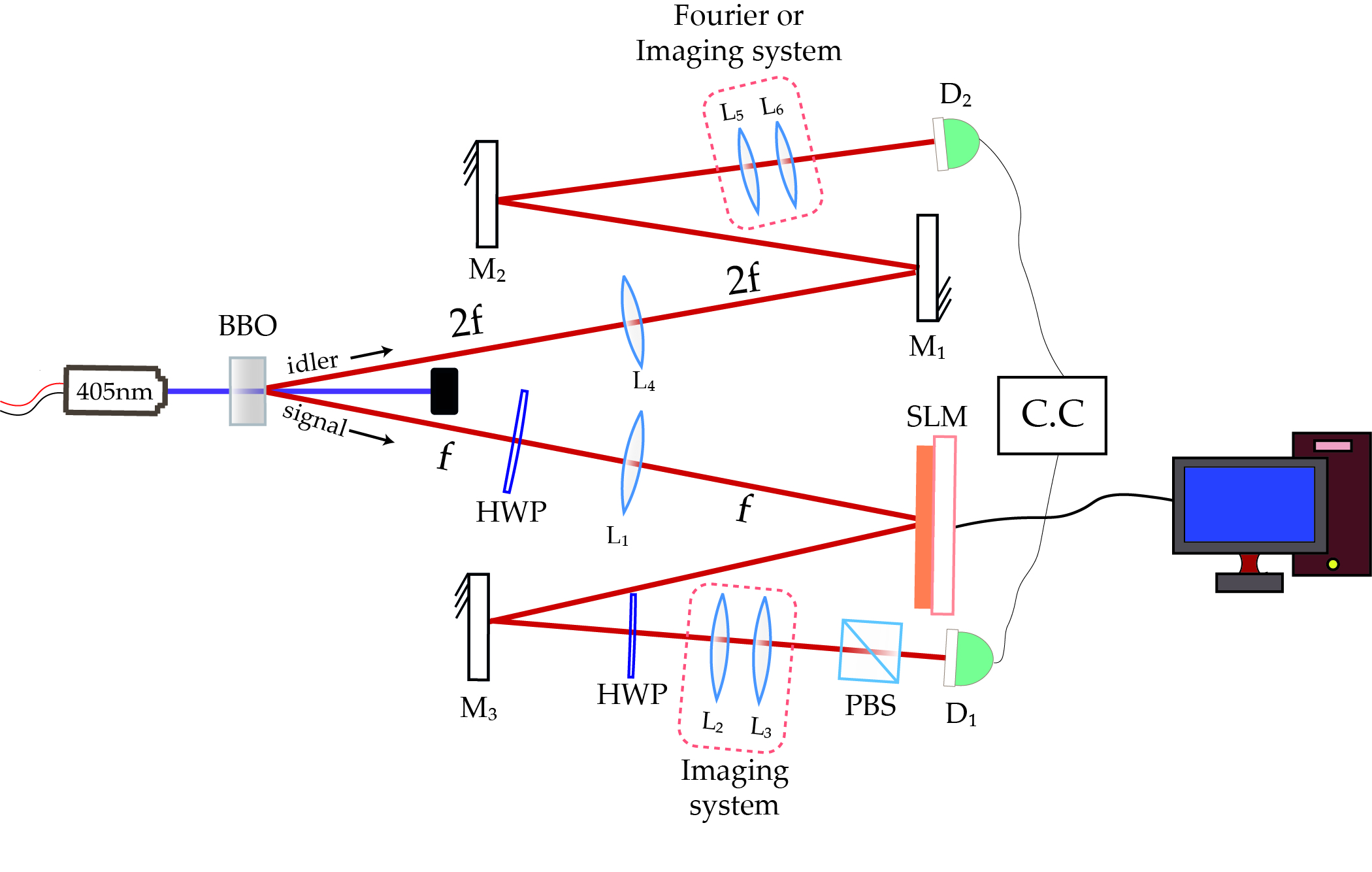}
	\caption{Experimental setup for optical processing with single photons: A laser of 405 nm in CW regime pumps a type-II nonlinear crystal. The 810 nm SPDC photons are selected using a frequency filter. The signal photon passes through a phase-to-amplitude modulation system composed of a sequence of optical elements; along this path, it experiences both a Fourier lens and an imaging system, and is detected at $D_1$. The idler photon passes through an imaging system followed by another system that can be configured either as a Fourier lens or as an imaging setup, and is detected at $D_2$.}
	\label{FIG1:experimental_setup}
\end{figure*}
Let us start by describing the experimental setup, illustrated in Figure \ref{FIG1:experimental_setup}. It is designed to investigate the integration of single-photon spatial encoding, using coincidence detection to perform the optical computation. In the process, the idler photon is detected and heralds a single-photon state, populating a superposition of states with some angular spectrum. This photon is incident on an SLM, and its wavefront is modulated, encoding a programmed binary function. The signal photon count rate is proportional to the integral of the binary function modulated on the SLM.

The laser source is a diode laser operating in continuous wave at 405 nm. It pumps a BBO nonlinear crystal cut for type I phase matching, producing SPDC photon pairs emitted in a non-collinear configuration and degenerate in frequency. Signal and idler photons are collected using bandpass interference filters centered at 810 nm with a 10 nm bandwidth.

The signal photon is sent through two optical systems: First, lens $L_1$ realizes the optical Fourier Transform of the field on the crystal plane onto the plane of the SLM, where phase modulation takes place. Second, the plane of the SLM is imaged onto $D_1$ detection plane.  The idler photon is sent through an imaging system from the crystal to mirror $M_1$. After reflection by $M_1$, the idler beam goes through an imaging or Fourier transform lens system, depending on the optical processing configuration selected. 

If imaging is selected, we have an overall imaging of the crystal plane onto the $D_2$ idler detection plane and an overall Fourier transform of the crystal plane onto the $D_1$ signal detection plane. In this case, there will be no correlation between the detection positions of signal and idler photons. However, if the Fourier transform lens system is chosen, there is an overall Fourier transform of the crystal plane onto detection planes $D_1$ signal and $D_2$ idler. In this case, there will be a correlation between the detection positions of signal and idler photons.   

Our goal is to analyze the performance of the optical processing scheme in these two configurations, spatially correlated and uncorrelated photon pairs. Both configurations impact the state preparation of the heralded single photon state and the performance of the processor.

\section{Two-photon quantum state and SPDC spatial correlations}
\label{sec:state}

 In this section, we calculate the coincidence counting rate for signal and idler photons for the two experimental configurations we have explored and mentioned above: spatially correlated and uncorrelated photon pairs.
 
The quantum state of the light emitted in spontaneous parametric down conversion is often approximated by \cite{walborn10}: 
\begin{equation}
\ket{\psi}=\int\hspace{-2mm}\int \hspace{-1mm} d\textbf{q}_{1} d\textbf{q}_{2}\ \Psi(\textbf{q}_{1},\textbf{q}_{2})\ket{\textbf{q}_{1}}\ket{\textbf{q}_{2}},
\label{eq:state}
\end{equation}
where $\ket{\textbf{q}}$ represents a single-photon state with transverse momentum $\textbf{q}$ and indices 1 and 2 refer to signal and idler modes, respectively. The function
\begin{equation}
\Psi(\textbf{q}_{1},\textbf{q}_{2}) = v(\textbf{q}_{1}+\textbf{q}_{2})\, \gamma(\textbf{q}_{1}-\textbf{q}_{2}),
\end{equation}
represents the biphoton joint amplitude, where $v(\textbf{q})$ denotes the normalized angular spectrum of the pump beam at the crystal's exit plane, and $\gamma(\textbf{q}_{1}-\textbf{q}_{2})$ is the phase-matching function.

The coincidence detection probability as a function of the transverse coordinates at the detection planes $\bm{\rho_1}$ and $\bm{\rho_2}$ is proportional to the function \cite{walborn10}
\begin{equation}
P(\bm{\rho}_1,\bm{\rho}_2) \propto \bra{\psi}\bm{E}_1^\dagger(\bm{\rho}_1)\bm{E}^\dagger_2(\bm{\rho}_2)\bm{E}_1(\bm{\rho}_1)\bm{E}_2(\bm{\rho}_2)\ket{\psi}, 
\end{equation}
where $\bm{E}(\bm{\rho})$ is the electric field operator at position $\bm{\rho}$.  
For weak pump beam it is usual to associate a wave function $\Psi$ to the two photon state so that $P(\bm{\rho}_1,\bm{\rho}_2) = |\Psi(\bm{\rho}_1,\bm{\rho}_2)|^2$, and 
\begin{equation}
\Psi(\bm{\rho}_1,\bm{\rho}_2) = \bra{0}\bm{E}_1(\bm{\rho}_1)\bm{E}_2(\bm{\rho}_2)\ket{\psi}.  
\label{psiu}
\end{equation}

This approach is intuitive and provides a field-like representation for the two-photon coincidence probability distribution.
The electric field operator is given by
\begin{equation}
\bm{E}_{1(2)}(\bm{\rho}_{1(2)})= \int d\bm{q}_{1(2)} \, \e^{-i\bm{q}_{1(2)}.\bm{\rho}_{1(2)}} \, 
 \bm{a}_{1(2)}({\bm{q}_{1(2)}}),
\label{operator}
\end{equation}
where the annihilation operator $ \bm{a}_{1(2)}({\bm{q}_{1(2)}})$ annihilates photons in the optical mode with transverse momentum component ${\bm{q}_{1(2)}}$.  
 
Using these operators in Eq. \eqref{psiu} we obtain 
the two-photon wave function: 
\begin{align}
\Psi(\bm{\rho}_1,\bm{\rho}_2) = & \int\hspace{-2mm}\int \hspace{-1mm} d\textbf{q}_{1} d\textbf{q}_{2} \, \Psi(\textbf{q}_{1}\,\textbf{q}_{2}) 
\,\e^{-i\textbf{q}_1 .\bm{\rho}_1}\, \e^{-i\textbf{q}_2 .\bm{\rho}_2}  ,
\label{eq:wf2}
\end{align}
If the crystal is thin enough so that the phase matching function is constant in the paraxial region $\gamma(\textbf{q}) \propto 1$, then the wave function is
\begin{align}
\Psi(\bm{\rho}_1,\bm{\rho}_2) = & \int\hspace{-2mm}\int \hspace{-1mm} d\textbf{q}_{1} d\textbf{q}_{2} \, v(\textbf{q}_{1} +\textbf{q}_{2}) 
\,\e^{-i\textbf{q}_1 .\bm{\rho}_1}\, \e^{-i\textbf{q}_2 .\bm{\rho}_2}.
\label{eq:wf3}
\end{align}

\subsection{COR and UNC configurations}

Let us now apply Eq.~\eqref{eq:wf3} to our experimental system in two different configurations called COR and UNC, whose meaning will become clear later. 

The COR configuration corresponds to the experimental scheme where the idler beam is imaged onto mirror $M_1$ and Fourier transformed from $M_1$ to $D_2$ detection plane, while the signal beam is Fourier transformed from the crystal plane to the SLM and then imaged to $D_1$ detection plane. In this case, the wave function is given by:
\begin{align}
\Psi(\bm{\rho}_1,\bm{\rho}_2) = &\int\hspace{-2mm}\int \hspace{-1mm} d\textbf{q}_{1} d\textbf{q}_{2} \, v(\textbf{q}_{1} +\textbf{q}_{2}) \, {\cal A} (\textbf{q}_{1}) \\ \nonumber &\times \, \delta(\textbf{q}_1 - \bm{\rho}_1) 
\, \delta(\textbf{q}_2 - \bm{\rho}_2).
\label{eq:wf4}
\end{align}

In this wavefunction, ${\cal A} (\textbf{q}_{1})$ represents the modulation by the SLM after the Fourier transforming operation implemented by $L_1$. $\delta(\textbf{q}_1 - \bm{\rho}_1)$ represents the mapping of the momentum $\textbf{q}_1$ at the source into position $\bm{\rho}_1$ at the SLM plane, which is imaged to $D_1$ detection plane. $\delta(\textbf{q}_2 - \bm{\rho}_2)$ represents the mapping of the momentum $\textbf{q}_2$ onto position $\bm{\rho}_2$ realized by lenses $L_5$ and $L_6$. 

Finally, we assume the plane wave approximation for the pump, which allows replacing $v(\textbf{q}_{1} +\textbf{q}_{2})$ with $\delta(\textbf{q}_{1} +\textbf{q}_{2})$. Performing the integrations in $\textbf{q}_{1}$ and $\textbf{q}_{2}$ we obtain:
\begin{equation}
\Psi(\bm{\rho}_1,\bm{\rho}_2) =  {\cal A}(\bm{\rho}_{1}) \,  \delta(\bm{\rho}_1 + \bm{\rho}_2).
\label{eq:wf5}
\end{equation}

The coincidence counting rate is:
\begin{align}
P(\bm{\rho}_1,\bm{\rho}_2) = \big|{\cal A}(\bm{\rho}_1) \,  \delta(\bm{\rho}_1 + \bm{\rho}_2)\big|^2.
\label{eq:wf6}
\end{align}

This result emphasizes the spatial correlations between signal and idler photons given by the delta function and justifies the choice of the term COR for this configuration. We can also observe the effect of the SLM modulation given by ${\cal A}(\bm{\rho}_1)$. 

For the UNC configuration, the wavefunction is given by: 
\begin{align}
\Psi(\bm{\rho}_1,\bm{\rho}_2) = &\int\hspace{-2mm}\int \hspace{-1mm} d\textbf{q}_{1} d\textbf{q}_{2} \, v(\textbf{q}_{1} +\textbf{q}_{2}) \, {\cal A} (\textbf{q}_{1}) \\ \nonumber &\times \, \delta(\textbf{q}_1 - \bm{\rho}_1) 
\, \e^{i\textbf{q}_2 . \bm{\rho}_2}.
\label{eq:wf7}
\end{align}

Notice that the only difference from the previous case is the presence of the term $\e^{i\textbf{q}_2 . \bm{\rho}_2}$ instead of $\delta(\textbf{q}_2 - \bm{\rho}_2)$. The reason is that lenses $L_5$ and $L_6$ now form an imaging system from mirror $M_1$ to detection plane $D_2$. As a result, the crystal plane is also imaged onto $D_2$, and the mapping between momentum and position variables no longer occurs.

Using the plane wave approximation for the pump angular spectrum again and performing the integrals, we get
\begin{equation}
\Psi(\bm{\rho}_1,\bm{\rho}_2) =  {\cal A}(\textbf{q}_{1}) \,  \e^{-i(\bm{\rho}_1 - \bm{\rho}_2)}\,,
\label{eq:wf8}
\end{equation}
and the coincidence count rate is
\begin{align}
P(\bm{\rho}_1,\bm{\rho}_2) = \big|{\cal A}(\bm{\rho}_1) \big|^2\,.
\label{eq:wf9}
\end{align}

This result shows that the coincidence count rate is independent of the position $\bm{\rho}_2$ of detector $D_2$, which justifies the term UNC. Nevertheless, the coincidence rate still depends on the SLM modulation function. 

It is also important to note that the temporal correlation between the signal and idler photons is always present; otherwise, coincidence detections would not be observed.

\section{Polarization assisted optical integrator}\label{sec3}

The phase-only SLM can be converted into an amplitude modulator for the purpose of binary encoding and reading. In order to realize this operation, we use the polarization of the incident light on the SLM and its technical characteristic that it only modulates the horizontal polarization component.

After modulating the horizontal component and not the vertical, we can project the polarization state onto the diagonal polarization basis. This causes interference between the modulated and unmodulated components, and the result ranges from zero to a maximum value depending on the modulation phase.

This operation can be described by considering the case of a light field incident on the SLM that is constant in the transverse spatial plane, approximately like a plane wave. The polarization state is prepared in the linear diagonal state with the aid of a half-wave plate. Therefore, the field right before the SLM is given by:
\begin{eqnarray}
    \vec{E}_{in}(x,y) = E_0 \frac{\hat{h} + \hat{v}}{\sqrt{2}}, 
\label{eq1}
\end{eqnarray}
where $E_0$ is constant and $\hat{h}$ ($\hat{v}$) is a unit vector in the horizontal (vertical) direction. 

The SLM modulates only the horizontal $\hat{h}$ component, and the field after modulation writes:
\begin{eqnarray}
    \vec{E}_{in}(x,y) = E_0 \frac{\hat{h} \e^{\ii f(x,y)} + \hat{v}}{\sqrt{2}}, 
    \label{eq2}
\end{eqnarray}
where $f(x,y)$ is the modulation function.

After the SLM, the field propagates through a second half-wave plate, which rotates $\hat{h}$($\hat{v}$) vector to diagonal (anti-diagonal) $\frac{\hat{h} + \hat{v}}{\sqrt{2}}$($\frac{\hat{h} - \hat{v}}{\sqrt{2}}$). The rotated field is given by:
\begin{eqnarray}
    \vec{E}(x,y) =\frac{ E_0}{\sqrt{2}}\{ [1+\e^{\ii f(x,y)}]\hat{h} - [1-\e^{\ii f(x,y)}]\hat{v}\}.
    \label{eq3}
\end{eqnarray}

The intensity of the $\hat{h}$ and $\hat{v}$ polarization components can be measured with a polarizing beam splitter. Moreover, in the present work, we are interested in measuring the whole transverse distribution using a bucket detector:
\begin{eqnarray}
    {\cal I}_{\hat{h}(\hat{v})} = |E_0|^2 \sqrt{2} \int_{\mathbb{R}^2}  (1\pm \cos [f(x,y)])\,dxdy. 
    \label{eq4}
\end{eqnarray}

If we choose the modulation function to be $f(x,y) = \arccos[g(x,y)]$, the quantity ${\cal T} = {\cal I}_{\hat{h}} - {\cal I}_{\hat{v}}$ gives the integral of $g(x,y)$:
\begin{eqnarray}
    {\cal T} \propto \int_{\mathbb{R}^2} g(x,y)\,dxdy. 
    \label{eq5}
\end{eqnarray}

In this section, we demonstrated that polarization can serve to convert phase modulations into amplitude modulations, enabling optical integration in a manner analogous to that described in Ref. \cite{lemos14}. It should be noted that this is a fully classical approach, for which the spatial correlations between the SPDC photon pairs are not yet relevant.

\section{DQC1 and spatial correlations}\label{sec4}
\begin{figure}[h]
\centering
	\includegraphics[width=0.45\textwidth]{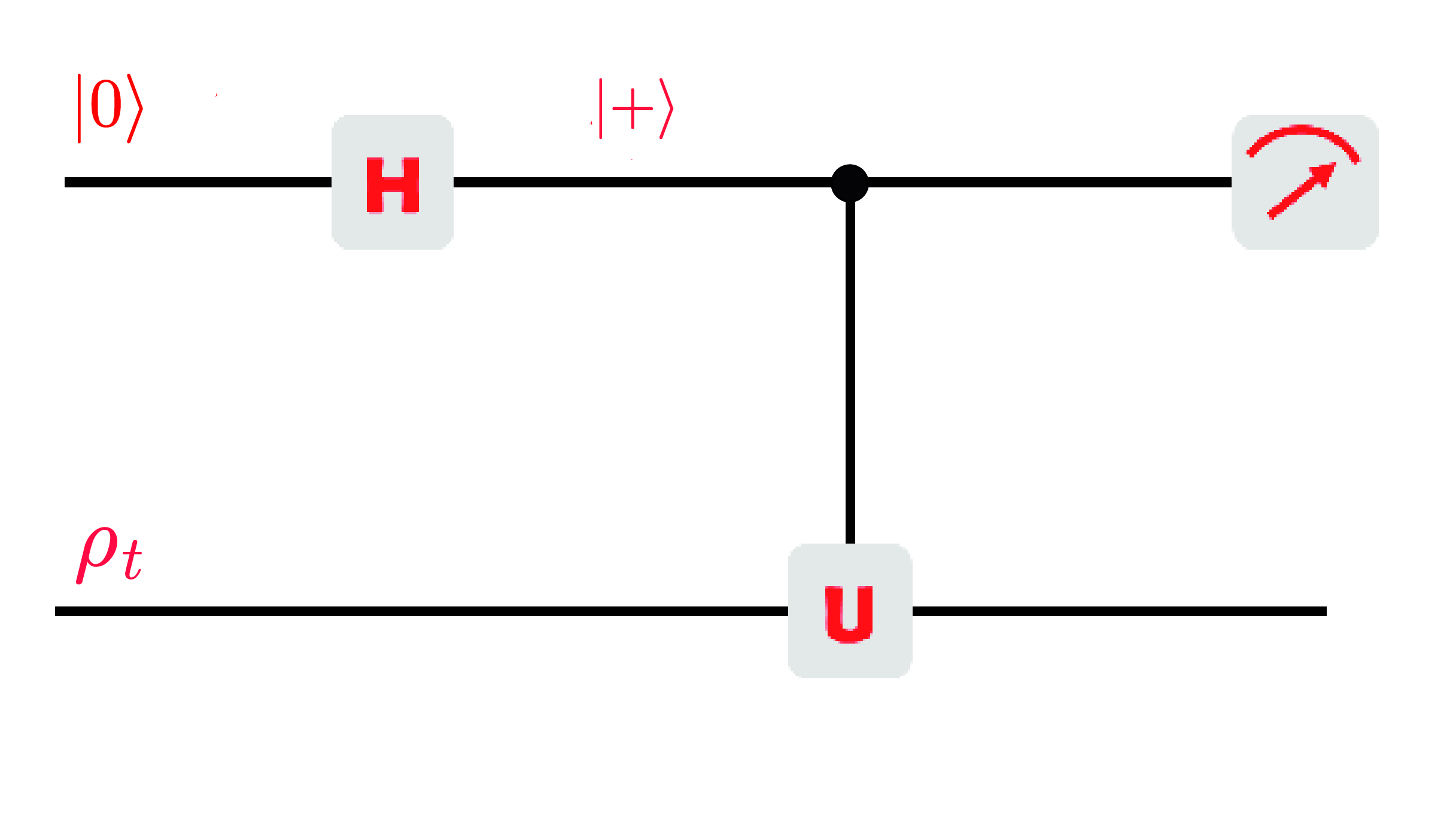}
	\caption{Quantum circuit diagram for Deterministic Quantum Computation with one qubit. The Hadamard gate $H$ acts on qubit $\ket{0}$ producing the state $\ket{+}=\frac{1}{\sqrt{2}}(\ket{0}+\ket{1})$, while the controlled operation $U$ acts on the system represented by density matrix $\rho_t$.}
	\label{DQC1}
   
\end{figure}

Deterministic quantum computation with one qubit (DQC1) (see Fig. \ref{DQC1}) is a method that allows the calculation of the trace of a normalized unitary matrix $U$. It only requires one qubit prepared as $\ket{+}=\frac{1}{\sqrt{2}}(\ket{0}+\ket{1})$ and a maximally mixed state $\rho_t$. The unitary is conditionally applied to $\rho_t$ and measuring the qubit provides the necessary information to calculate the trace of $U$.

In fact, Ref. \cite{meyll15} has shown that implementing DQC1 using single-photon polarization to encode the qubit and spatial light modulation to perform the unitary $U$ is the quantum analogue of classical optical integration with an SLM \cite{lemos14}. Therefore, the present experimental scheme can be interpreted both as a quantum optical integrator and as a photonic realization of DQC1. In the following, we summarize the main ideas presented in Ref. \cite{meyll15}.

We begin with a single photon with diagonal polarization representing the qubit $\ket{+}$. The transverse mode of the photon should be evenly distributed over the SLM plane. In mathematical terms, it compares to the maximally mixed state
\begin{equation}
    \rho_t = C \sum_{i,j} \ket{i, j} \bra{i, j},
\end{equation}
where $\ket{i,j}$ represents a single-photon state with transverse mode localized at the pixel position $(i,j)$, and $C$ is a normalization constant.

A controlled operation $S$ is implemented on the tranverse degrees of freedom using the SLM, which is capable of modulating only the horizontal component. We can describe such an operation as
\begin{equation}
    S = \ket{H}\bra{H} \otimes U + \ket{V}\bra{V} \otimes \mathbb{1},
\end{equation}
in which $\ket{H}$ and $\ket{V}$ are the horizontal and vertical polarization states, respectively, and
\begin{eqnarray}
    U = \sum_{i,j}\e^{-\ii\phi_{ij}} \ket{i, j} \bra{i, j},
\end{eqnarray}
where $\phi_{ij}$ is the phase applied in the SLM's cell $(i,j)$. Note that $U$ is a diagonal matrix with trace $\sum_{i,j} \e^{-\ii\phi_{ij}}$.

The polarization and spatial degrees of freedom of the modulated light are entangled, such that the reduced density matrix of the qubit, which is obtained by tracing out the spatial degree of freedom, is given by \cite{meyll15}:
\begin{equation}
\rho = \frac{1}{2}
\begin{pmatrix}
1 & C \sum_{i,j} \e^{-\ii\phi_{ij}} \\
C \sum_{i,j} \e^{\ii\phi_{ij}} & 1 
\end{pmatrix}\,.
\label{eq6}
\end{equation}

It is important to note that the sum of all SLM phase factors appears in the off-diagonal terms of the density matrix. Consequently, the expectation values of the Pauli operators $\sigma_x$ and $\sigma_y$ for this state are given by $\langle \sigma_x\rangle= C \sum_{i,j} \cos \phi_{ij}$ and $\langle \sigma_y\rangle = C \sum_{i,j} \sin \phi_{ij}$. Therefore, measuring both $\sigma_x$ and $\sigma_y$ therefor provides direct access to $Tr(U)=\sum_{i,j} \e^{-\ii\phi_{ij}}$.

For optical integration based on binary encoding, the phase $\phi_{ij}$ can only take the values $0$ and $\pi$, as discussed in the next section. In this case, $\langle \sigma_y\rangle=0$, while $\langle \sigma_x\rangle$ is proportional to the difference between the number of white pixels ($\phi=0$) and black pixels ($\phi=\pi$). The integral of the binary function can then be retrieved using the expression $N/2\times(1+\langle\sigma_x\rangle)$, where $N$ is the total number of pixels.

In summary, DQC1 computes the trace of a normalized unitary matrix through measurements on a single-photon state. This represents the quantum and discrete analogue of the optical integration described in the previous section: quantum, because it involves single-photon states; and discrete, because it operates on discrete matrices.

An important remark here is that all other degrees of freedom of the photon are traced out, and only the polarization state is measured.

\section{Result and Discussion}\label{sec11}
\subsection{Integration of a binary function}

\begin{figure}[H]
    \centering
    \includegraphics[width=0.55\linewidth]{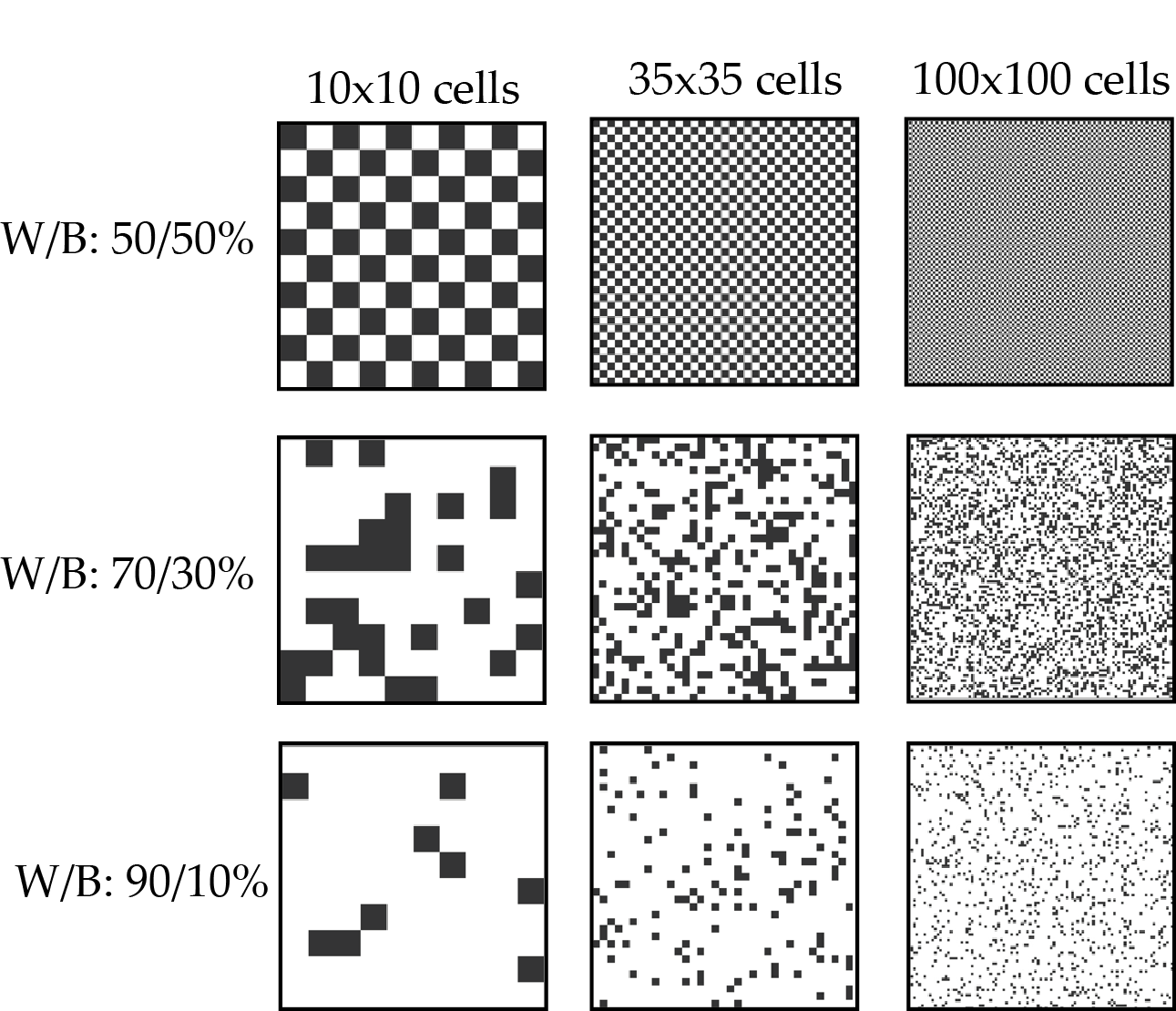}
    \caption{Binary maps prepared in the SLM mask. White cells are groups of pixels of the SLM modulating with phase $\phi = 0$. Black cells are groups of pixels of the SLM modulating with phase $\phi = \pi$. }
    \label{pannel}
\end{figure}

We analyze the optical integration with heralded single photons and binary functions. The SLM is programmed with binary masks, meaning that ``white'' (W) cells apply phase $0$ and ``black'' (B) cells apply phase $\pi$. Therefore, the light reflected by a W cell is transmitted through the polarizing beam splitter (PBS) before the signal detector, and the light reflected by B cells is reflected by the PBS. The percentage of W and B cells is varied: 50/50\%, 30/70\% and 10/90\%. For each percentage, the number of cells in the matrices is also varied. The SLM masks are displayed in Fig. \ref{pannel}.

\begin{figure}[H]
    \centering
    \includegraphics[width=0.55\linewidth]{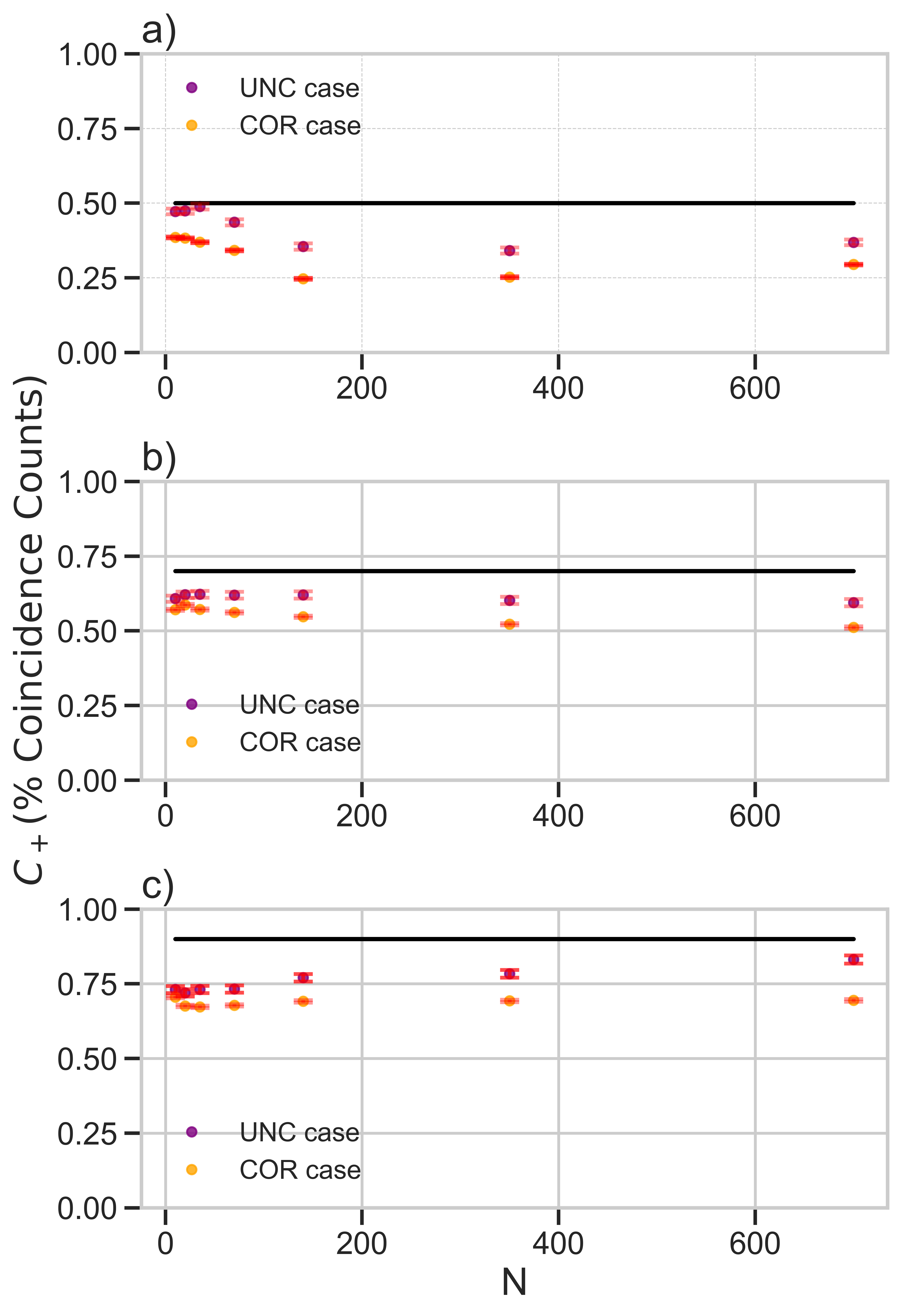}
    \caption{Percentage coincidence counting rate $C_+$, with error bars due to poissonian statistics, as a function of $N$, where $N$ is the dimension of the modulated square matrix $N \times N$ on the mask generated by the SLM (some examples are illustrated in Fig. \ref{pannel}). Each set of measurements was taken with different proportions of white and black blocks: a) 50\%/50\%, b) 70\%/30\%, and c) 90\%/10\%.}
    \label{fig:graph_proportion}
\end{figure}

Fig.~\ref{fig:graph_proportion} shows the measurement results. $C^+(\%) = C^+/(C^+ + C^-)$, where $C^+$ and $C^-$ are the coincidence rates at the transmission and reflection outputs of the PBS, respectively. In other words, the measured coincidence rate is equal to the integral of the binary function, expressed as a percentage of W cells for $C^+(\%)$ and B cells for $C^-(\%)$.

The measurements are performed in two different arrangements. When the optical system $L_5$/$L_6$ implements imaging of mirror $M_1$ onto $D_2$ detector in the idler beam, we have no spatial correlations between signal and idler photons (UNC). From the perspective of the heralded single photon, the absence of correlations means that the transverse spatial degrees of freedom are traced out (circles). When $L_5$/$L_6$ implements a Fourier Transform, we have spatially correlated signal and idler photons (COR). This means that the spatial degrees of freedom of the heralded single photon are not traced out (squares).

The three plots show that the results for the UNC case approach the percentage imprinted in the SLM, represented by the black line, better than the COR case. This means that the integration works better with UNC photons than with COR ones. At first sight, this result may seem surprising. However, this is expected when we analyze it from the DQC1 perspective: the UNC configuration traces out the spatial degrees of freedom and fits the DQC1 operation conditions. 

\subsection{Spatial sampling capability}

In a second set of measurements, we evaluate the ability of the single-photon wave packet to sample the spatial modulation at the SLM surface. The goal is to realize the remote state preparation of the heralded photon so that it is as transversely broad as possible, allowing a single photon to interact with the largest possible surface on the SLM. In this experiment, we do not use a matrix, but only a one-dimensional encoding. The $D_1$ detector is kept and in bulked detection, while $D_2$ is displaced along the vertical direction with single slit (with aperture $\sim0.3$ mm), while the SLM is prepared with a binary mask consisting of alternating W and B stripes of 70 pixels each ($\sim$0.56\,mm).

\begin{figure}[H]
    \centering
    \includegraphics[width=0.75\linewidth]{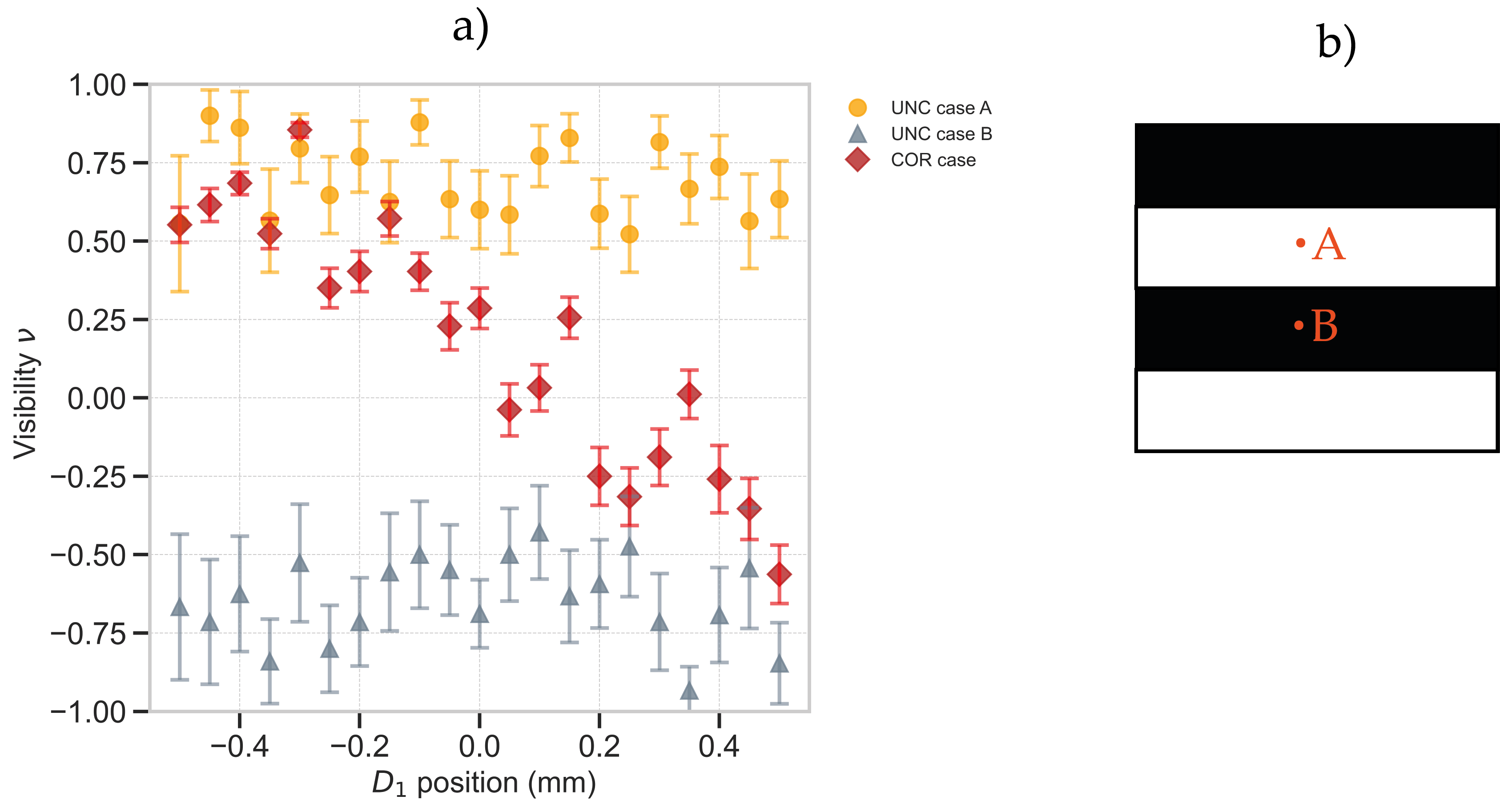}
    \caption{Coincidence visibility as a function of $D_1$ displacement for three measurement sets. The UNC cases A and B are measured by initially centering detector $D_1$ (imaging plane) at exemplified points A and B of the modulated mask shown in b).}
    \label{fig:enter-label}
\end{figure}

Fig.~\ref{fig:enter-label} shows the measurement results. The visibility is given by $v = (C^+ - C^-)/(C^+ + C^-)$, where $C^+$ ($C^-$) is the coincidence counting rate at the W (B) output of the PBS. Negative visibility simply means that $C^- > C^+$. The results show that using the UNC configuration leads to nearly constant visibility as a function of the $D_2$ vertical position. Each plot (circles and diamonds) corresponds to a different $D_1$ vertical position. For the COR configuration, the visibility changes from a maximum positive to a maximum negative when $D_2$ is scanned (triangles).

We interpret the results as an indication that the UNC configuration better approaches the goal of producing a broad heralded single photon than the COR configuration. The reason is that the change from positive to negative visibility shows that both W and B stripes interact with the photon wave packet on the SLM, while in the COR case, the wave packet only interacts with W in one instance or B in another. This result is also consistent with the better performance of the UNC case for the integration scheme presented previously.

We observe that the UNC regime performs better in capturing the global features of the modulated mask, even through detection region (limited by the slit in $D_1$) is smaller than that individual stripes and even more to the length of the total modulated field. This behavior is evidenced by the relatively constant visibility observed in $D_1$ displaced. In the ideal case, it would be possible to infer the ratio between white and black regions of the mask from a single coincidence event. In contrast, under the COR regime, the system is able to capture the local features; however, the amount of information that can be extracted is limited by the size of the detection region.

\section{Conclusion}\label{sec13}

In conclusion, we have performed an experimental investigation concerning the use of heralded single photons for optical processing through the transverse spatial degrees of freedom of light. We present two theoretical approaches that can be used to describe the system and compare the performance for two types of heralding of the single photon state. In the COR configuration, the spatial correlations between the gating photon and the heralded one are preserved, while in the UNC configuration, they are canceled. Our results indicate that the UNC configuration is more useful for reading the information programmed on the SLM surface and performing the integration operation. These results are consistent with the interpretation of the scheme as a DQC1 algorithm, where the trace over the spatial degrees of freedom of the photonic qubit is performed. These are the first steps toward optical processing in the quantum regime and using single-photon states to process large amounts of information encoded in the transverse spatial degrees of freedom of light.

\section*{Author Contributions}
LMF, RCSP, MHM conducted the experiment and analysed the results; EID  contributed to the theory of experiment; RMG and RMA discussed the results;  PHSR conceived and conducted the experiment, contributed to the theoretical framework, and discussed the results. All authors contributed writing and reviewing the manuscript.
\section*{Funding}
This work has been supported by the following Brazilian research agencies: Conselho Nacional de Desenvolvimento Cient\'{\i}fico e Tecnol\'ogico (CNPq - DOI 501100003593), Coordena\c c\~{a}o de Aperfei\c coamento de Pessoal de N\'\i vel Superior (CAPES DOI 501100002322), Funda\c c\~{a}o de Amparo \`{a} Pesquisa do Estado de Santa Catarina (FAPESC - DOI 501100005667), Instituto Nacional de Ci\^encia e Tecnologia de Infraestruturas Qu\^antica e
Nano para Aplica\c c\^oes Convergentes INCT-IQNano 406636/2022-2, Instituto Nacional de Ciência e Tecnologia em Informação Quântica, and Funda\c c\~{a}o de Amparo \`{a} Pesquisa e Inova\c{c}\~{a}o do Estado de Goi\'{a}s
(FAPEG). EID acknowledges
further support from CNPq under Grant No.
409673/2022-6.

\section*{Data Availability}
No datasets were generated or analyzed during the
current study.
\section*{Declaration}
\textbf{Conflict of Interest}: The authors declare no competing interests.

\bibliography{sn-bibliography}

\end{document}